# Dependence of the Spin Transfer Torque Switching Current Density on the Exchange Stiffness Constant

Chun-Yeol You[1]

[1]Department of Physics, Inha University, Incheon 402-751, Korea

We investigate the dependence of the switching current density on the exchange stiffness constant in the spin transfer torque magnetic tunneling junction structure with micromagnetic simulations. Since the widely accepted analytic expression of the switching current density is based on the macro-spin model, there is no dependence of the exchange stiffness constant. When the switching is occurred, however, the spin configuration forms C-, S-type, or complicated domain structures. Since the spin configuration is determined by the shape anisotropy and the exchange stiffness constant, the switching current density is very sensitive on their variations. It implies that there are more rooms for the optimization of the switching current density with by controlling the exchange stiffness constant, which is determined by composition and the detail fabrication processes.

The magnetic random access memory (MRAM) with the spin transfer torque (STT)[1,2] is one of the promising spintronic devices. For the successful commercialization of STT-MRAM, the understanding of the underlying physics of the switching mechanism by the STT is essential. Many theoretical studies have addressed about the physical origin and the bias dependence of the in-plane and out-of-plane STT using simple free electron models,[1,2] first principle calculations,[3] and Keldysh non-equilibrium Green's function methods.[4,5,6,7,8] The simple picture of the current induced magnetization switching (CIMS) is anti-damping process due to the in-plane STT term, which is generated by the deposition of the angular momenta of the conduction spins. When the anti-damping term satisfy the un-stability condition, the switching is occurred.[9,10,11,12] Based on un-stability conditions, the analytic expression of the switching current density, $J_c$, can be derived for the macro-spin model. Since the most of the switching conditions studies have been investigated based on the macro-spin model, there is no contribution of the exchange stiffness constant, $A_{ex}$. However, the importance and the role of $A_{ex}$ in the STT excited spin wave have been already addressed in prior literatures.[10,13,14,15,16,17,18,19] Even though $A_{ex}$ is considered in the expression for $J_c$ in the prior studies, none of them clearly show the detail relations between $J_c$ and $A_{ex}$. Rippard et al.[13] and Slonczewski[14] pointed out that the contribution of $A_{ex}$ in the excited in the spin wave energy, but not related with $J_c$. Zhou et al.[15] performed micromagnetic simulations with two extremes, unrealistically large exchange stiffness ($A_{ex}$ = 30x10$^{11}$ J/m) and realistic exchange stiffness ($A_{ex}$ = 3x10$^{11}$ J/m), which is bulk value of Co. They claimed that incoherent and non-zero $k$ spin waves can be excited and it alters the switching boundary conditions. However, they missed more realistic range of the $A_{ex}$ (≤ $3 \times 10^{11}$ J/m), and detail dependence of $J_c$. Sato et al.[16] focused on the junction size effect on $J_c$ and the thermal stability by introducing nucleation size, which is related with the domain wall width and $A_{ex}$. Sun et al.[17,18] reported the size dependence of $J_c$ and thermal stability, and claimed that the exchange length is important in the switching mechanism and thermal stability. Berkov et al.[19] studied the detail of STT driven spin dynamics, but they paid their attention to the precessional motion of the spin. Even though there are many prior studies about the STT driven spin dynamics, no one explicitly pointed out the dependence of $J_c$ on $A_{ex}$, and expect large variation of $J_c$ with small changes of $A_{ex}$.

In this study, we investigate systematically the switching current density, $J_c$, dependence on the exchange stiffness constant, $A_{ex}$, for the typical magnetic tunneling junction (MTJ) structures with wide range of $A_{ex}$, which are realistic. We employ micromagnetic simulator OOMMF (Object Oriented MicroMagnetic Framework)[20] with public STT extension module.[21] Surprising, it has been revealed that the switching current densities are very sensitive function of the $A_{ex}$. We find that even small change of $A_{ex}$ leads large variation of switching current density, which is not easy to expect from macro-spin model. When the switching occurs, the detail spin configurations are far from single domain, they form C- type, S- type, or more complex domain structures. Furthermore, the lateral dimension of the MTJ determines the possible excitation wave vector of the spin wave and shape anisotropy energy, the $A_{ex}$ dependences become complex.

Furthermore, $A_{ex}$ is fundamental physical quantities of the ferromagnetic materials, and it is closely related with Curie temperature and the saturation magnetization. In addition, the spin polarization, which is important in spintronic devices, is also related with the $A_{ex}$. For example, the CoFeB alloy is widely used materials for the free layer of the STT-MRAM. Recently, we have found that $A_{ex}$ of the CoFeB alloy are varied with the fabrication conditions, such as Ar gas pressure, substrates, seed layers, compositions, and annealing conditions.[22] Such variation of $A_{ex}$ is understandable, because $A_{ex}$ is sensitive function of the distance between magnetic atoms and number of nearest neighborhood.[23] Since the composition of the B is varied with the annealing processes, $A_{ex}$ is also changed with the detail fabrication conditions. Therefore, the study of $J_c$ dependence on $A_{ex}$ will provide additional recipes for the further reduction of $J_c$.

We consider typical STT-MRAM structure with an exchange biased SyF (Synthetic Ferrimagnet) layer ($F_3$/NM/$F_2$).[24] The saturation magnetization $M_s$ and thicknesses of $F_{1, 2, 3}$ layers are $1.3 \times 10^6$ A/m and 2 nm, respectively. The thicknesses of NM and insulator layers are 1 nm. The cross-section of the nano-pillar is an ellipse of 60×40 and 80×40nm$^2$, with the cell size of 1×1×1 nm$^3$. No crystalline anisotropy energy is considered in this study for the simplicity. The exchange stiffness constant $A_{ex}$ varies from 0.5 to $3.0 \times 10^{-11}$ J/m, and the Gilbert damping constants $\alpha$ is fixed to 0.02. The exchange bias field of $4 \times 10^5$ A/m is assigned to the long axis of the ellipse (+$x$-direction) for the $F_3$ layer. For the simplicity, we consider only in-plane STT and ignore the out-of-plane STT contributions. More details of micromagnetic simulations can be found elsewhere.[21]

We calculate $J_c$ of the 60×40 nm$^2$ and 80×40 nm$^2$ ellipses with various $A_{ex}$. We varies $A_{ex}$ from 0.5 to $3.0 \times 10^{-11}$ J/m, because the exchange stiffness constants of typical ferromagnetic materials such as Co, Fe, Ni, and NiFe are 3.0, 2.1, 0.9 and $1.3 \times 10^{-11}$ J/m, respectively. Furthermore, $A_{ex}$ values have been reported for $Co_{80}Fe_{20}$ and $Co_{72}Fe_{18}B_{10}$ are 2.61~$2.84 \times 10^{-11}$ J/m,[25] and our measurements results for $Co_{40}Fe_{40}B_{20}$ are scattered from 1.0 to $1.4 \times 10^{-11}$ J/m depends on the detail fabrication processes.[22]

Figure 1 shows the results of $J_c$ of P to AP and AP to P switching for various $A_{ex}$ for $60\times40$ nm$^2$ and $80\times40$ nm$^2$ ellipses. Surprisingly enough, the data are somewhat scattered 2.46 ~ 2.7×10$^{11}$ A/m$^2$, 2.65 ~ 3.73×10$^{11}$ A/m$^2$, and 3.11 ~ 5.16×10$^{11}$ A/m$^2$ for each switching. The difference between minimum and maximum of $J_c$ are 9.8, 40.7, and 65.9 % for P to AP and AP to P ($60\times40$ nm$^2$) and P to AP ($80\times40$ nm$^2$) switching processes, respectively. We find that the variation of $J_c$ is more pronounce in the elongated ellipse. For the $80\times40$ nm$^2$ ellipse, the higher $J_c$ (~5×10$^{11}$ A/m$^2$) is found for $A_{ex}$ of 1.2 ~ 1.7×10$^{-11}$ J/m and similar $J_c$ (~3.5×10$^{11}$ A/m$^2$) are obtained for $A_{ex}$ <1.2×10$^{-11}$, > 1.8×10$^{-11}$ J/m, despite of different $A_{ex}$. As shown in Fig. 1, the variation of $J_c$ is clear as a function of $A_{ex}$, but it seems the dependences are somewhat complicate, and it is not easy to explain the $J_c$ variations with simple model.

According to the widely accepted analytic expression for $J_c$, these numerical results are striking and must be addressed. If we include the exchange energy term in the macro-spin model, the $J_c$ is read as:[10,13,14]

$$J_c \sim \frac{\alpha}{a_1}\left[H_{eff} + \frac{1}{2}(N_y + N_z - 2N_x)M_s + \frac{2A_{ex}}{\mu_0 M_s}k_x^2\right]. \quad (1)$$

Here, Where $N_{x,y,z}$ are the demagnetization factors of the switching layer ($N_z \gg N_y \geq N_x$ for thin ellipse), and $H_{eff}$ is the effective field including an external, perpendicular STT term, stray, and Oersted fields. $\alpha$ and $a_1$ is the Gilbert damping constant and $a_1 = \eta_p \frac{\hbar}{2e\mu_0 M_s d_s}$. Here, $\eta_p$, $d_s$, $M_s$, $\mu_0$, and $\hbar$ are the spin polarization of the polarizer layer, the thickness of the switching layer, the saturation magnetization, and reduced Plank's constant, respectively. The wave vector $k_x$ is zero for the macro-spin model.

According to the Eq. (1), the relation between $J_c$ and $A_{ex}$ is straightforward, however, our numerical results cannot be explained by the simple analytic expression. Because Eq. (1) is derived with the infinite plane assumption, however, our numerical simulations have been done for the finite lateral size, so that the $k_x$ is limited due to the boundary conditions. Due to the strong demagnetization field of the nano-structure, the $k_x$ is weakly quantized by the size of the structure. There meaning of the weakly quantization will be discussed later.

In order to reveal the origin of the $J_c$ dependence on $A_{ex}$, we show the snap shots of each case in Figs. 2 ~6. We select the snap shots to represents the switching mechanisms. The movie files of each switching are available in the Supplementary materials (Animations 1~5 are corresponding to the Figs. 2~6).[26] First, we show the snap shots of 60×40 nm$^2$ (AP to P switching) for $A_{ex}$ of 1.0×10$^{-11}$ J/m, where we find the lowest $J_c$ in Fig. 1. Fig. 2 (a) and (b) show the formation of the multiple domains, where the domain size is order of 20~30 nm. The switching is occurred after formation of somewhat complicate domain structures. The highest $J_c$ is found for $A_{ex}$ (=2.9×10$^{-11}$ J/m), and the snap shots are shown in Fig. 3 (a)~(d). Due to the larger $A_{ex}$, the variation of the spin configuration is slower, and no complex domain structures are found. Thicker wall is also found in Fig. 3 (a), and the C-type domain is formed as shown in Fig. 3 (b). The switching is occurred without complex domain structures. Therefore, we can conjecture that the spin configurations of the switching mode are fairly different for different $A_{ex}$. Furthermore, the spin configurations are function of not only $A_{ex}$, but also the shape anisotropy, which determine the size of the domain and domain wall with $A_{ex}$.

We also investigate different size of the 80×40 nm$^2$ ellipse (P to AP switching) as shown in Fig. 1. In this case, the highest $J_c$ is found for $A_{ex}$ of 1.2 ~ 1.7×10$^{-11}$ J/m, and either larger or smaller $A_{ex}$ give lower $J_c$. We show the snap shots of $A_{ex}$ = 0.5, 1.5, and 3.0×10$^{-11}$ J/m in Fig. 4~6. For $A_{ex}$ of 0.5×10$^{-11}$ J/m, multiple domains are formed before the switching (see Fig. 4 (a)~(d)), and the spin configurations are varying very rapidly. The other limit, $A_{ex}$ of 3.0×10$^{-11}$ J/m, it shows different behavior. Figure 6 (a)~(b) show well defined domain wall between two domains at each side. The domain wall widths are comparable to the domain size. Even when the switching is occurred, the spin configurations are slowly varying. Therefore, the detail switching mechanism is quite different from $A_{ex}$ of 0.5×10$^{-11}$ J/m case in Fig. 4. Let us see the Fig. 5 (a)~(b) for $A_{ex}$ =1.5×10$^{-11}$ J/m, where the highest $J_c$ is found. Compare to the Figs. 6 ($A_{ex}$=3.0×10$^{-11}$ J/m), the wall width is much thinner. And the domain wall is much stable when the two side domains are fluctuating. Therefore, the stable domain wall must be removed to achieve switching, it requires higher $J_c$. The animations of the switching process in the Supplementary materials are useful to obtain more insights and the different switching modes.

So far, we show the $J_c$ is sensitive function of the $A_{ex}$. During the switching processes, C-, S-type, or complex domains are formed, which cannot be properly treated in the macro-spin model. And the domain and domain wall structures are determined by the $A_{ex}$ and the shape anisotropy. Since the exchange energy contribution in the excited spin wave is $\frac{2A_{ex}}{\mu_0 M_s}k_x^2$, and $k_x$ is not a continuous variable in the nano-structure. As shown in Fig. 2~6, the domain and domain wall size will limit the value of $k_x$, we call this is weak quantization of $k_x$. Since the domain and domain wall sizes are coupled with the shape of the nano-structure and $A_{ex}$, $\frac{2A_{ex}}{\mu_0 M_s}k_x^2$ term must be a complex function of the shape of nano-structure and $A_{ex}$, not a simple linear function of $A_{ex}$. It must be noted that we do not discuss the details of about 60×40 nm$^2$ ellipse (P to AP switching) case. The variations of $J_c$ of this case are weaker than others, and they show noticeable different behaviors from the AP to P switching case. In these simulations, we assume the same $a_1$ for the simplicity, the main differences between two switching processes are stray fields from synthetic antiferromagnetic layer structure.[21] It implies the stray field plays an important role in the formation of the domain structures.

In conclusions, we find the switching current density $J_c$ is a function of the exchange stiffness constant $A_{ex}$. Surprisingly enough, the variation is 65.9 % for 80×40 nm$^2$ ellipse, and the dependence is not a simple. Since the exchange contribution in the spin wave term will be determined by the coupling of the shape anisotropy and exchange energy through the weakly confined spin wave vector $k_x$. Based on our finding, there is more room to reduce the switching current density by engineering the exchange stiffness constant and shape of the ellipse.

**Acknowledgements**

This work is supported by the NRF funds (Grand No. 2010-0023798 and 2010-0022040) by Ministry of Education, Science, and Technology of Korea.

[18] J. Z. Sun, R. P. Robertazzi, J. Nowak, P. L. Trouilloud, G. Hu, D. W. Abraham, M. C. Gaidis, S. L. Brown, E. J. O'Sullivan, W. J. Gallagher, and D. C. Worledge, Phys. Rev. B, **84**, 064413 (2011)..

[19] D. V. Berkov and J. Miltat, J. Mag. Mag. Mater. **320**, 1238 (2008).

[20] M. J. Donahue and D. G. Porter, OOMMF User's Guide, Ver. 1.0, Interagency Report NISTIR 6376, NIST, USA (1999).

[21] C.-Y. You, cond-mat/arXiv:1201.4707

[22] J. Cho, J. Jung, K.-E. Kim, S. Lee, and C.-Y. You, Bull. of the Kor. Phys. Soc. Meeting, Daejon, P2-D126, Apr. 26, (2012).

[23] C. Kittel, "Introduction to Solid State Physics", 8$^{th}$ Ed. Wiley, p. 323 (2005).

[24] M.-H. Jung, S. Park, C.-Y. You, and S. Yuasa, Phys. Rev. B **81**, 134419 (2010): C.-Y. You, J. Yoon, S. –Y. Park, S. Yuasa, M. –H. Jung, Cur. Appl. Phys. **11**, e92 (2011).

[25] C. Bilzer, T. Devolder, P. Crozat, and C. Chappert, J. Appl. Phys. **101**, 074505 (2007).

[26] The corresponding animations are available from [URL will be inserted by AIP].
10

**Figure Captions**

Fig. 1 Switching current density as a function of the exchange stiffness constant $A_{ex}$. 60×40 nm$^2$ ellipse for P to AP (red circle) and AP to P (green circle) switching. 80×40 nm$^2$ ellipse for P to AP (blue circle) switching.

Fig. 2 Snap shots of the 60×40 nm$^2$ ellipse AP- to P-state switching processes with $A_{ex}$ of 1.0×10$^{-11}$ J/m at $t =$ (a) 6.3, (b) 8.6, (c) 8.9, and (d) 9.3 ns.

Fig. 3 Snap shots of the 60×40 nm$^2$ ellipse AP- to P-state switching processes with $A_{ex}$ of 2.9×10$^{-11}$ J/m at $t =$ (a) 4.2, (b) 4.3, (c) 4.8, and (d) 5.0 ns.

Fig. 4 Snap shots of the 80×40 nm$^2$ ellipse P- to AP-state switching processes with $A_{ex}$ of 0.5×10$^{-11}$ J/m at $t =$ (a) 7.9, (b) 8.2, (c) 9.3, and (d) 9.9 ns.

Fig. 5 Snap shots of the 80×40 nm$^2$ ellipse P- to AP-state switching processes with $A_{ex}$ of 1.5×10$^{-11}$ J/m at $t =$ (a) 4.0, (b) 5.2, (c) 7.4, and (d) 9.0 ns.

Fig. 6 Snap shots of the 80×40 nm$^2$ ellipse P- to AP-state switching processes with $A_{ex}$ of 3.0×10$^{-11}$ J/m at $t =$ (a) 4.0, (b) 7.0, (c) 9.3, and (d) 9.5 ns.

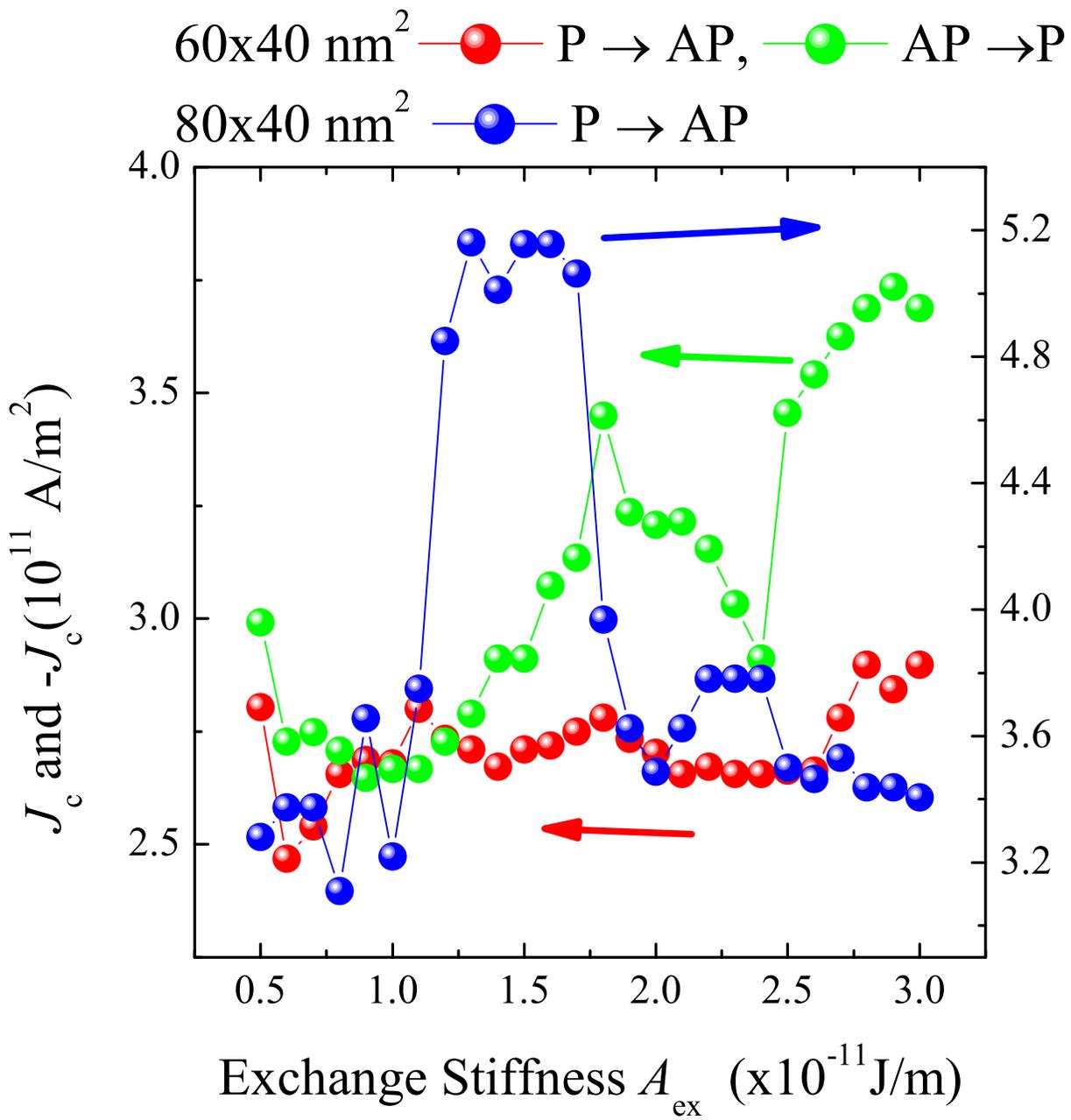

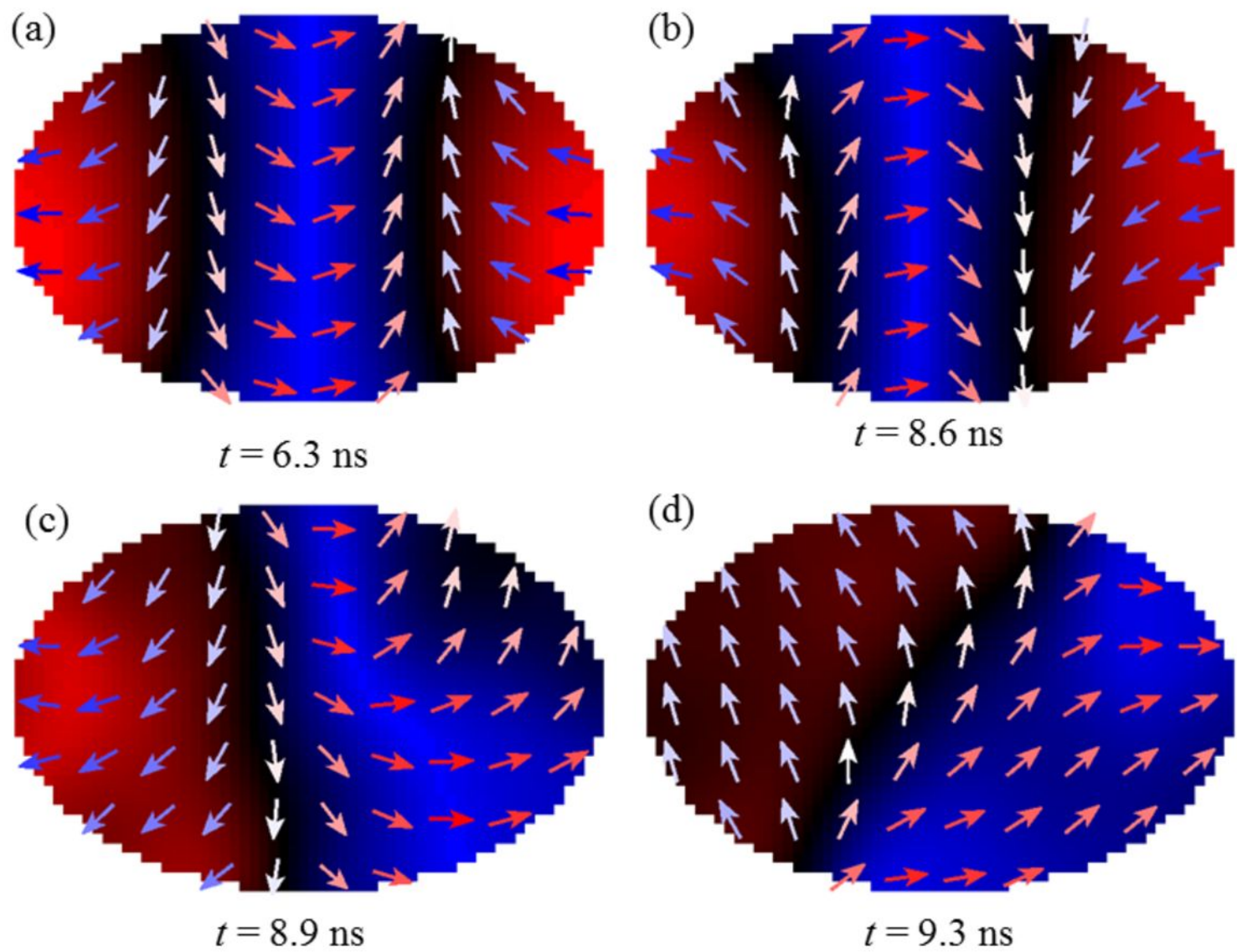

(a) $t = 6.3$ ns

(b) $t = 8.6$ ns

(c) $t = 8.9$ ns

(d) $t = 9.3$ ns

Fig. 2

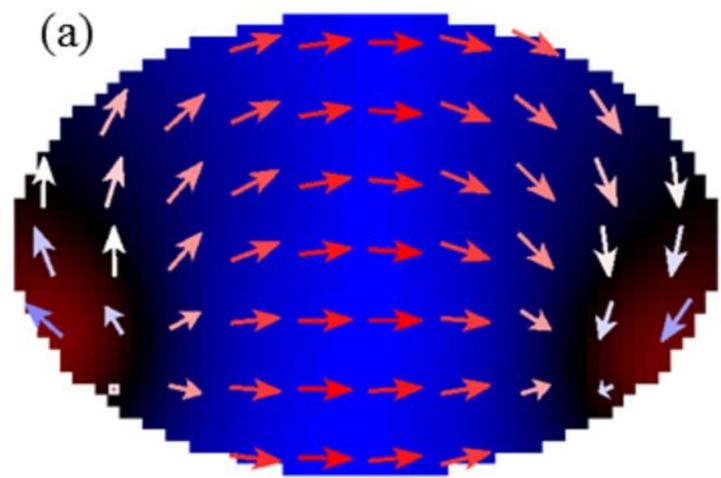 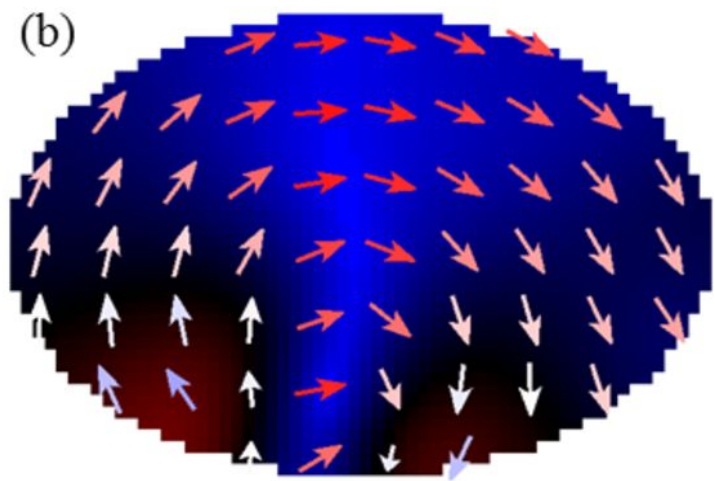

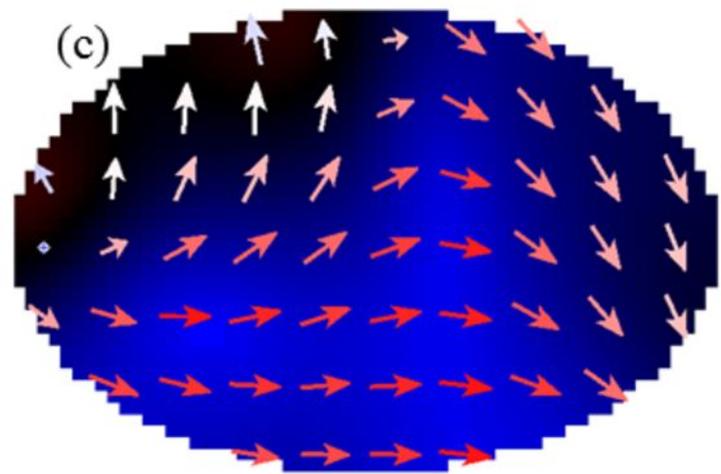 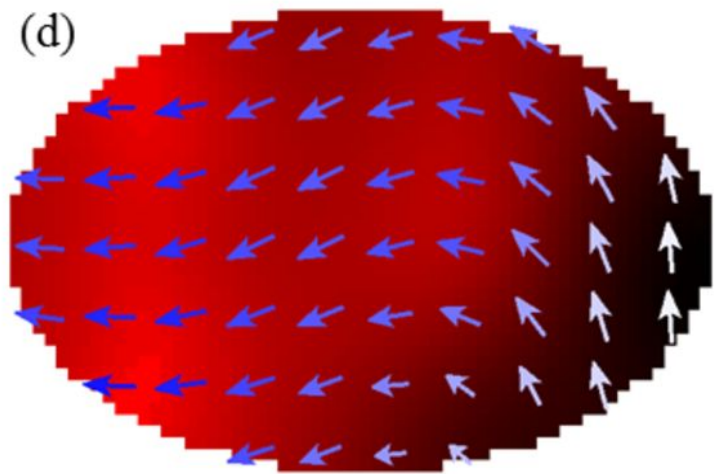

Fig. 3

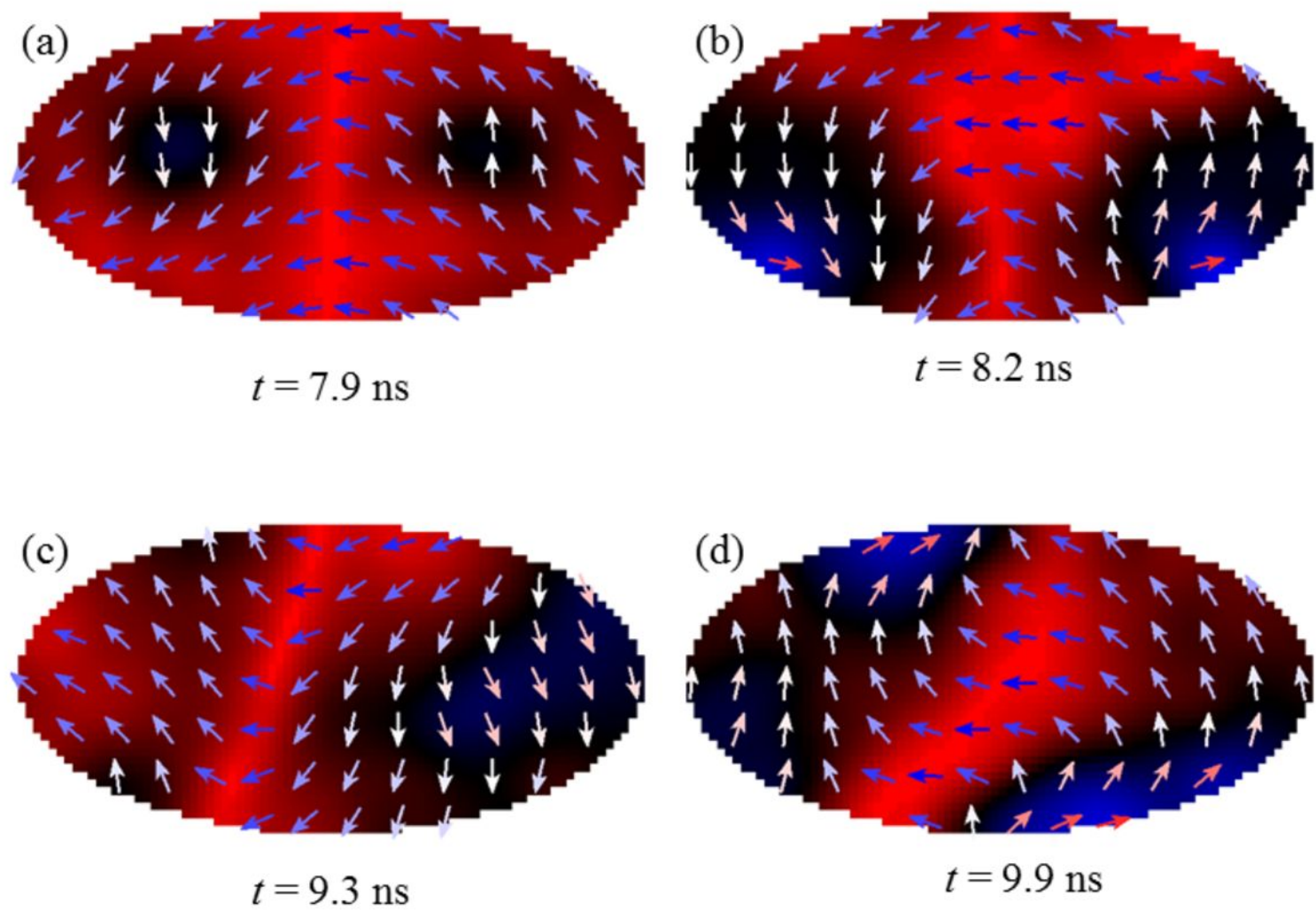

Fig. 4

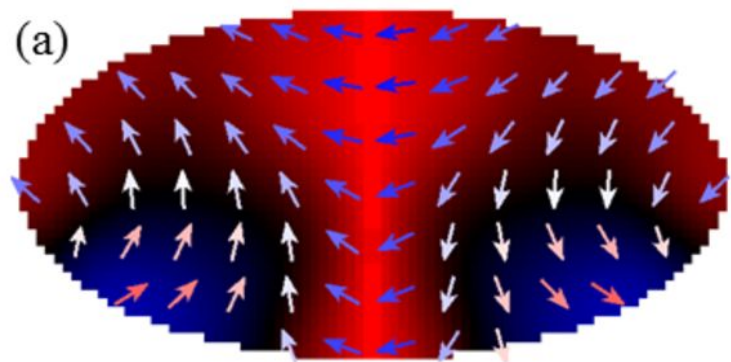 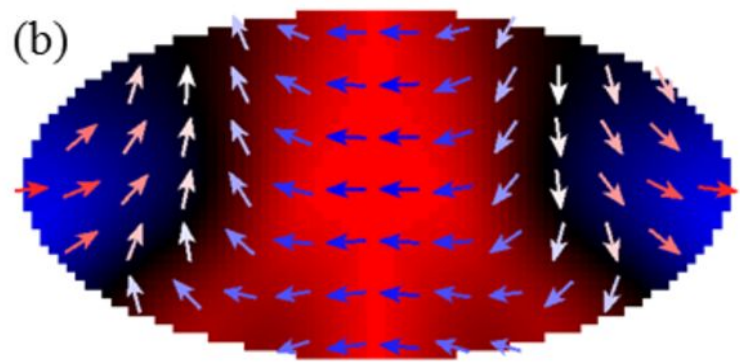

(a) t = 4.0 ns

(b) t = 5.2 ns

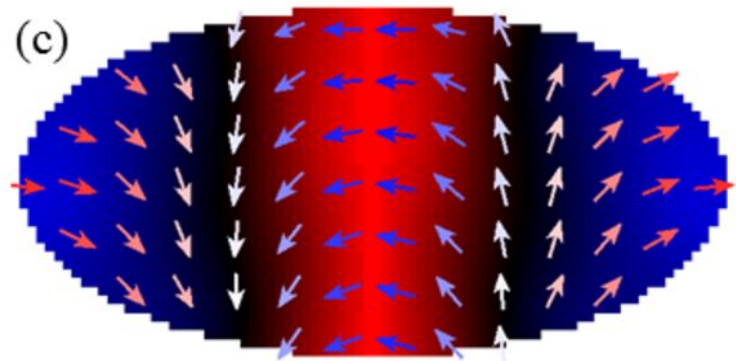 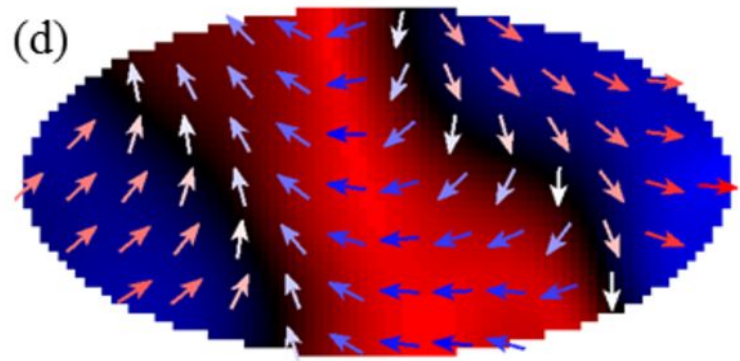

(c) t = 7.4 ns

(d) t = 9.0 ns

Fig. 5

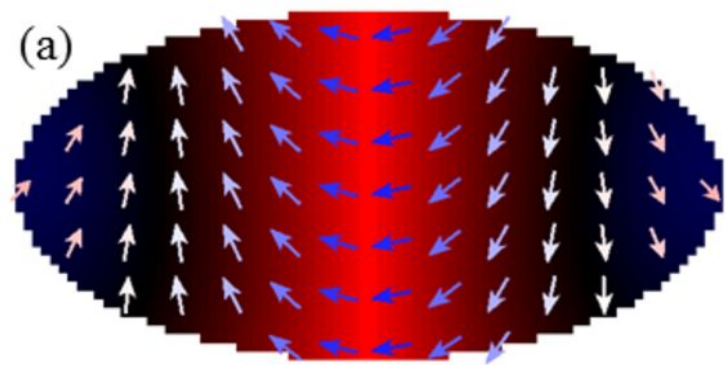
(a) t = 4.0 ns

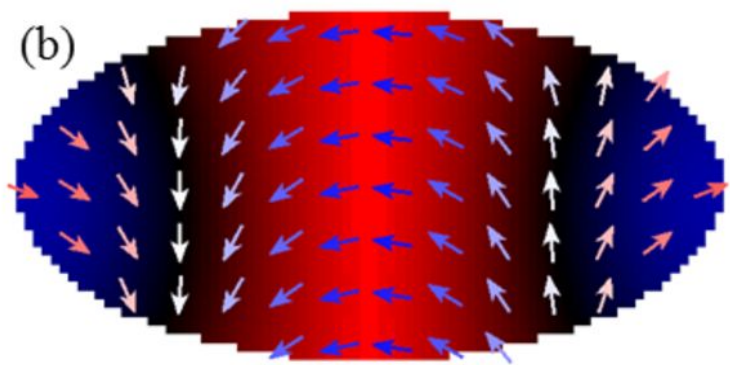
(b) t = 7.0 ns

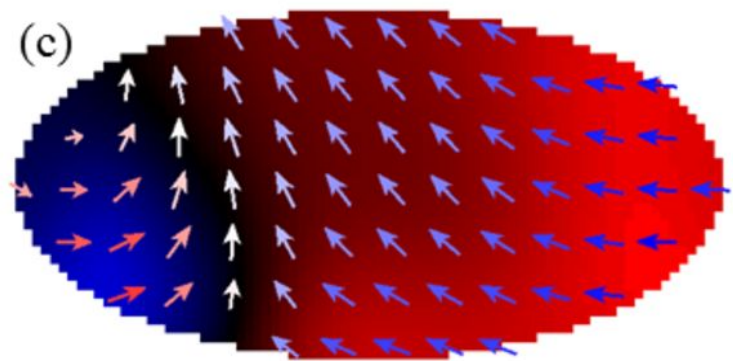
(c) t = 9.3 ns

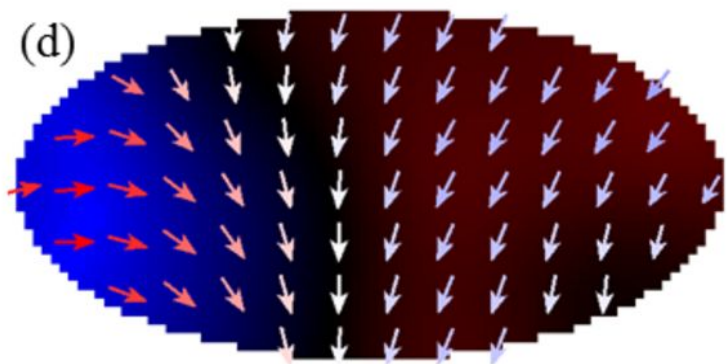
(d) t = 9.5 ns

Fig. 6